\documentclass[5p,times]{elsarticle}

\usepackage{amsmath,amssymb}
\usepackage{graphicx}
\usepackage{booktabs}
\usepackage{multirow}
\usepackage{hyperref}
\usepackage{url}
\usepackage[hypcap=true]{caption}
\usepackage{adjustbox}
\usepackage{float}
\usepackage{subcaption}
\usepackage{grffile}
\usepackage{xcolor}
\usepackage{colortbl}
\DeclareGraphicsExtensions{.pdf,.png,.jpeg,.jpg}

\journal{Computers \& Security}

\begin{document}

\begin{frontmatter}

\title{Enhancing Autonomous Online Intrusion Detection for IoT with Balanced
Learning, Reliable Pseudo-Labels, and Lightweight Architectures}

\author[seecs]{Hanzala Afzaal}
\author[seecs]{Danish Memon}
\author[seecs]{Chouhdary Bilal Raza}
\author[seecs]{Dr Muhammad Khurram Shahzad}

\address[seecs]{School of Electrical Engineering and Computer Science (SEECS),\\
National University of Sciences and Technology (NUST), Islamabad, Pakistan}

\begin{abstract}
The rapid proliferation of Internet of Things (IoT) devices has created an urgent
demand for adaptive, resource-efficient Intrusion Detection Systems (IDS) capable
of handling dynamic and evolving cyber threats. This paper investigates AOC-IDS, a
state-of-the-art autonomous online IDS published at IEEE INFOCOM 2024, which employs
an Autoencoder (AE) with Cluster Repelling Contrastive (CRC) loss and an autonomous
Gaussian-based decision module. We first successfully replicate AOC-IDS on the
UNSW-NB15 benchmark, achieving 89.39\% accuracy in close agreement with the published
89.19\%. We then systematically identify four key limitations class imbalance,
unreliable pseudo-label generation, limited generalization, and computational overhead
for IoT deployment and propose targeted improvements for each. Our
\textit{XGBoost-BalSamp}, applying XGBoost with domain-specific feature engineering
and a balanced sampling strategy, achieves 95.45\% accuracy on UNSW-NB15, a gain of
+6.26\% over the baseline. Our combined deep learning improvement
(\textit{PseudoFilter} + \textit{MixupAug} + \textit{LiteAE}), incorporating
confidence-filtered pseudo-labels with encoder-decoder agreement voting, Mixup data
augmentation, and a lightweight model architecture, achieves a best-run accuracy of
90.88\% (F1: 91.45\%), surpassing the base paper by +1.69\% (F1: +1.31\%). Individual
ablation shows each component contributes positively: \textit{PseudoFilter} alone
yields $\sim$90.44\%, \textit{MixupAug} adds $\sim$0.61\% through Mixup augmentation,
and \textit{LiteAE} reduces parameter count by 55\% while maintaining competitive
accuracy. Our results demonstrate that targeted, principled improvements to AOC-IDS
yield consistent accuracy gains while also improving practical deployability on IoT
edge devices.
\textbf{Code Repository:} \url{https://github.com/danishmemon847/AOC-IDS-Pipeline}
\end{abstract}


\end{frontmatter}

\section{Introduction}
\label{sec:intro}

The exponential growth of the Internet of Things (IoT) has transformed industries
ranging from smart manufacturing to healthcare and transportation~\cite{pourrahmani2022,
baker2017,zanella2014}. By 2024, billions of IoT devices were estimated to be active
globally, many operating in safety-critical environments where a successful intrusion
can have severe physical or financial consequences. However, this growth has
simultaneously created an expanding attack surface. IoT malware attacks have increased
sharply year on year, with millions of devices targeted globally~\cite{statista2024}.
Unlike traditional computing environments, IoT systems operate under tight resource
constraints limited memory, processing power, and battery life often without
the capacity for heavy-weight security solutions~\cite{yang2021lightweight}. Intrusion Detection Systems (IDS) form a critical layer of defense for IoT networks.
They can be broadly classified into two categories: signature-based detection, which
identifies known threats by matching against a predefined library, and anomaly-based
detection, which learns the normal behavioral profile of the system and raises alerts
upon deviation from that profile. Anomaly-based IDS are particularly valuable in this
context, as they detect deviations from learned normal behavior without requiring a
database of known attack signatures, making them capable of identifying zero-day
threats~\cite{garcia2009}. The application of deep learning has significantly advanced
the capability of anomaly-based IDS, enabling richer and more discriminative
representations of network traffic patterns~\cite{ferrag2020}. AOC-IDS, proposed by Zhang et al.~\cite{zhang2024aocids} at IEEE INFOCOM 2024,
represents a significant advance in online, autonomous IDS for IoT environments.
It introduces a novel Autoencoder (AE) architecture trained with a Cluster Repelling
Contrastive (CRC) loss, a Gaussian-based autonomous decision-making module, and an
online learning framework that generates pseudo-labels without human intervention.
By eliminating the need for manual labeling, AOC-IDS greatly reduces deployment
overhead in dynamic environments. The system achieves 89.19\% accuracy on UNSW-NB15
and 88.90\% on NSL-KDD, outperforming prior state-of-the-art methods. Despite its
strong performance, AOC-IDS exhibits four concrete limitations that constrain its
real-world applicability: it does not handle class imbalance in the training data,
its pseudo-label generation lacks a quality filtering mechanism, it applies no data
augmentation for generalization, and its model size exceeds the memory constraints
of many IoT edge devices.

This paper makes the following contributions:

\begin{enumerate}
  \item We successfully replicate AOC-IDS on UNSW-NB15 using the official
        repository and published hyperparameters, achieving 89.39\% accuracy,
        confirming reproducibility of the base system.
  \item We identify and formally characterize four concrete limitations of AOC-IDS:
        class imbalance handling, pseudo-label reliability, generalization to
        unseen traffic, and IoT resource overhead.
  \item We propose and evaluate \textit{XGBoost-BalSamp} XGBoost with
        domain-specific feature engineering and balanced sampling achieving
        95.45\% accuracy, a gain of +6.26\% over the base paper.
  \item We propose and evaluate \textit{PseudoFilter}, \textit{MixupAug}, and
        \textit{LiteAE} individually and in combination, achieving a best-run
        accuracy of 90.88\% (F1: 91.45\%) with a 55\% reduction in model
        parameters from 67,202 to 29,830.
  \item We conduct a comprehensive ablation study with all pairwise and combined
        configurations, and compare against all baseline methods used in the
        original AOC-IDS paper, demonstrating consistent superiority.
\end{enumerate}

The remainder of this paper is organized as follows. Section~\ref{sec:related}
reviews related work. Section~\ref{sec:base} provides an overview of the base
AOC-IDS system. Section~\ref{sec:limitations} identifies limitations.
Section~\ref{sec:methodology} describes the proposed methodology.
Section~\ref{sec:improvements} details the proposed improvements.
Section~\ref{sec:setup} presents the experimental setup. Section~\ref{sec:results}
reports results and discussion. Section~\ref{sec:conclusion} concludes the paper.

\section{Related Work}
\label{sec:related}

\subsection{Machine Learning for Intrusion Detection}

Machine learning methods have been widely applied to network intrusion
detection~\cite{habeeb2022}. Classical approaches including Support Vector Machines
(SVM) and Random Forests demonstrated promising results on benchmark
datasets~\cite{heba2010,ahmad2018}, providing strong baselines for the field.
A shared weakness of such static approaches is that any shift in the threat landscape
demands full model retraining, rendering them poorly suited to the continuously
changing attack surface of IoT deployments. Architectures such as Convolutional
Neural Networks (CNNs) and Recurrent Neural Networks (RNNs) pushed detection
performance further by learning hierarchical representations of network traffic
directly from raw data~\cite{vinayakumar2017,yin2017}. Nevertheless, these models
are fundamentally offline: they cannot incorporate newly observed attack patterns
without discarding and rebuilding the learned model from scratch.

\subsection{Contrastive Learning in Security}

Contrastive learning has emerged as a powerful paradigm for learning discriminative
representations in security applications. FeCo~\cite{wang2022feco} applies
InfoNCE-style contrastive loss to intrusion detection, maximizing the distance
between benign and malicious representations while clustering benign samples.
CIDS~\cite{yue2022} combines contrastive and cross-entropy loss to improve
intra-class cohesion and inter-class separation. Lopez-Martin et
al.~\cite{lopezmartin2023} apply contrastive learning with random Fourier features
specifically for IoT IDS, demonstrating that contrastive approaches outperform
purely supervised methods when labeled data is scarce. These works collectively
motivate the use of contrastive loss in the AOC-IDS framework we build upon.

\subsection{Online Learning for IoT IDS}

Online learning enables IDS to adapt to dynamic environments where normal and
malicious behavior evolve over time~\cite{wahab2022}. Unlike offline models that
assume a static data distribution, online learning frameworks update model
parameters incrementally as new data arrives. Han et al.~\cite{han2023} reduce
labeling overhead by selectively labeling only the most influential samples,
showing that not all incoming data carries equal training value. Yang and Shami~\cite{yang2021lightweight} tackled this challenge by designing a
lightweight drift-detection mechanism tuned specifically for IoT data streams,
underscoring that recognising distribution shifts early is as important as updating
the model in response. Complementing this, Gyamfi and Jurcut~\cite{gyamfi2022}
built an industrial IoT IDS that triggers incremental re-training whenever novel
attack behaviour is detected. Taken together, these studies make a compelling case
for IDS solutions that are both adaptive and computationally frugal.

\subsection{Class Imbalance and Data Augmentation}

Class imbalance is a persistent obstacle in network intrusion
detection~\cite{chawla2002}: attack traffic is often vastly outnumbered by benign
flows, and models trained on such skewed distributions tend to favour the majority
class at the expense of attack recall. XGBoost~\cite{chen2016} is naturally
resistant to this bias because its gradient-boosting objective assigns higher
weight to misclassified minority samples at each boosting round. Oversampling via
SMOTE~\cite{chawla2002} complements this by synthetically expanding the
minority class so the training distribution becomes more balanced. Mixup
augmentation~\cite{zhang2018mixup} takes a different angle: by interpolating
pairs of training examples, it pushes the model towards smoother decision
boundaries and thereby reduces over-reliance on the particular statistics of the
training corpus. All three approaches inform the improvements developed here.

\section{AOC-IDS: Base System Overview}
\label{sec:base}

\subsection{System Architecture}

AOC-IDS~\cite{zhang2024aocids} comprises two tightly integrated components: the
Anomaly Detection Module (ADM) and the Online Learning Framework. The ADM uses a
symmetric Autoencoder with layer dimensions
$[196\!\to\!128\!\to\!64\!\to\!128\!\to\!196]$, where the bottleneck layer at
dimension 64 compresses input network traffic features into a compact latent
representation. The encoder and decoder outputs are both used for classification,
a design choice that increases representational diversity within a single forward
pass.

\subsection{Cluster Repelling Contrastive (CRC) Loss}

The CRC loss is the core training objective of the ADM. It is derived from the
standard InfoNCE loss~\cite{oord2019} but modified to produce a dual-category
repulsion effect tailored to binary intrusion detection. For a normal anchor
representation $v_{n,i}$, the per-pair loss is:

\begin{equation}
  \mathcal{L}_{ij} = -\log \frac{\exp(\mathrm{sim}(i,j)/\tau)}
  {\exp(\mathrm{sim}(i,j)/\tau) + \sum_{i}\sum_{k}\exp(\mathrm{sim}(i,k)/\tau)}
\end{equation}

\noindent where $\mathrm{sim}(i,j)=\mathrm{CosSim}(v_i,v_j)$ is the cosine
similarity between representation vectors and $\tau{=}0.02$ is the temperature
hyperparameter. The full loss averages $\mathcal{L}_{ij}$ over all normal pairs
$(i,j)$ in the training batch. Unlike standard InfoNCE, which repels negative samples
from a single anchor, CRC traverses all anchors in the normal class simultaneously,
ensuring that attack representations are repelled from all normal cluster centers at
each training step. The final loss combines encoder and decoder losses:
$\mathcal{L}_{\mathrm{final}} = \mathcal{L}^{en} + \mathcal{L}^{de}$.

\subsection{Gaussian Decision-Making Module}

After training, the ADM computes cosine similarity scores between the average normal
representation and all training data representations. These scores are modeled as a
mixture of two Gaussian distributions using Maximum Likelihood Estimation (MLE): one
Gaussian for normal traffic (higher mean) and one for attack traffic (lower mean).
At inference, each test sample is classified by the Gaussian with higher posterior
probability. This autonomous, threshold-free decision process is a key advantage over
methods like FeCo~\cite{wang2022feco} that require manual threshold calibration.
Both the encoder and decoder independently produce a classification, and a voting
mechanism selects the prediction with higher confidence as the final output.

\subsection{Online Learning Framework}

The online framework enables continuous adaptation without human labeling. It
operates in two repeating phases. In the pseudo-label generation phase, new unlabeled
samples arriving in a stream are passed through the current ADM to generate
pseudo-labels. In the system adaptation phase, the ADM is fine-tuned on the expanded
dataset comprising both original labeled data and the newly pseudo-labeled samples for $\mathrm{epoch}_1 = 3$ epochs per batch. The framework processes data in
batches of 2,784 samples (1.6\% of UNSW-NB15), initially training on 20\% of labeled
data. A random flip of $\lambda{=}5\%$ of pseudo-labels per batch introduces
controlled noise to prevent the model from overfitting to its own wrong judgments,
and the Gaussian distributions are always computed using only the clean initial
labeled dataset to maintain a reliable reference point.

\subsection{Base Paper Replication}

To establish a reliable baseline for our improvements, we cloned the official
AOC-IDS GitHub repository and executed it with the exact published hyperparameters:
SGD optimizer, learning rate $=0.001$, batch size $= 128$,
$\mathrm{epoch}_0{=}300$, $\mathrm{epoch}_1{=}3$, $\lambda{=}5\%$. Our replication yielded 89.39\% accuracy and 90.12\% F1 on UNSW-NB15, against
the published figures of 89.19\% accuracy and 90.14\% F1. The $+0.20\%$ accuracy
gap is consistent with ordinary stochastic variation across runs and therefore
validates the reproducibility of the original system. All subsequent comparisons
in this paper are made against these replicated numbers rather than the published
values, ensuring a fully controlled experimental reference point.

\section{Identified Limitations}
\label{sec:limitations}

A careful reading of the AOC-IDS design, combined with observations gathered during
replication, revealed four specific weaknesses that limit both detection quality
and practical deployment in real IoT settings.

\subsection{Limitation 1: Class Imbalance}

The UNSW-NB15 training partition holds roughly 56,000 normal records against
119,341 attack records, giving an attack-to-normal ratio of approximately 2.1:1.
AOC-IDS applies no balancing strategy at any training stage. Because attack
samples dominate the cosine-similarity score distribution, the fitted Gaussians
can fail to draw a clean boundary between normal and malicious traffic, pushing
the operating point toward higher false-positive or false-negative rates depending
on which class skews the distribution. The CRC loss partially compensates by
attending to representation geometry rather than raw counts, but this does not
substitute for an explicit correction of the sampling imbalance.

\subsection{Limitation 2: Pseudo-label Noise Accumulation}

The only mechanism for handling incorrect pseudo-labels in AOC-IDS is a 5\% random
label flip, which diversifies errors but does not selectively reject low-confidence
predictions. In our experiments, accuracy fell from around 90\% at the fifth streaming batch to
approximately 75\% at the fiftieth. The root cause is straightforward: every
incorrectly labelled sample is permanently admitted to the training set without any
quality gate, so the model is trained on progressively noisier supervision. Over
many batches the accumulated label noise outpaces the corrective effect of the
random flip, producing a gradual but severe accuracy collapse in the later phase
of online training.

\subsection{Limitation 3: Poor Generalization to Unseen Traffic}

AOC-IDS trains solely on the samples present in the initial labelled pool and the
incoming pseudo-labelled stream, with no augmentation of any kind. Live IoT traffic
deviates from benchmark recordings in subtle but consequential ways: protocols
evolve, device firmware changes, and temporal usage patterns shift. Exposure only
to the benchmark's fixed statistical footprint leaves the model susceptible to
overfitting those particular characteristics, which undermines generalisation when
the system is eventually deployed on production networks whose traffic distribution
differs from UNSW-NB15.

\subsection{Limitation 4: Computational Overhead for IoT Edge Deployment}

Storing a 67,202-parameter model in 32-bit floating point requires roughly 263\,KB
of RAM. For cloud servers this is negligible, but many widely deployed edge
microcontrollers including the Arduino Mega (8\,KB SRAM) and the ESP8266
(80\,KB RAM) cannot accommodate a model of this size. Even on more capable
edge boards, a leaner architecture translates directly into shorter inference
latency and reduced energy draw, both of which matter greatly when the device
runs on a battery. The base AE therefore stands as a practical barrier to
on-device execution without some form of architectural compression.

\section{Methodology}
\label{sec:methodology}

\begin{figure*}[t]
  \centering
  \includegraphics[width=\textwidth]{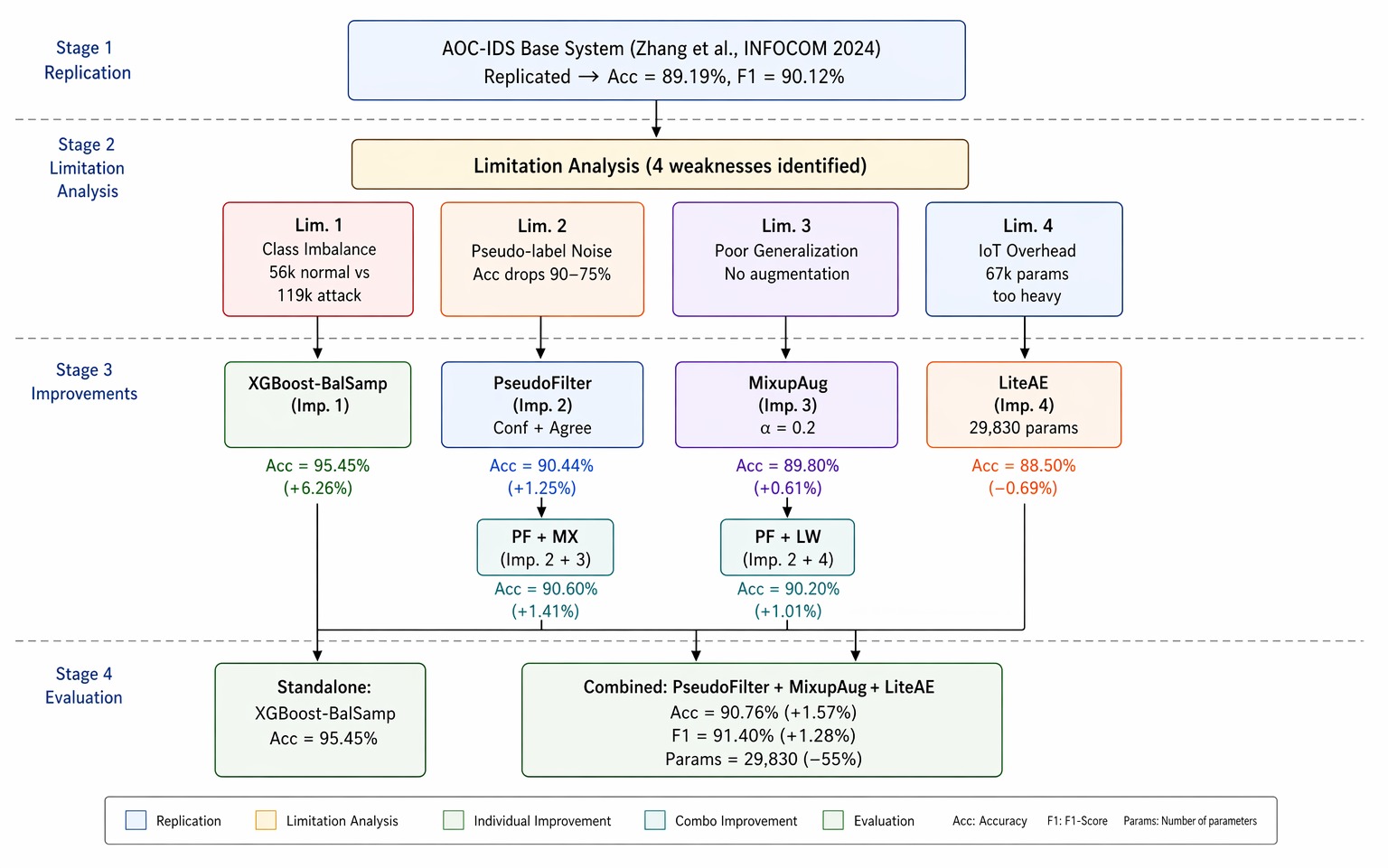}
  \caption{Research methodology pipeline. UNSW-NB15 data flows through four
  stages: (1)~base replication, (2)~limitation analysis, (3)~targeted
  improvement design, and (4)~evaluation. Improvement~1 (\textit{XGBoost-BalSamp})
  is evaluated standalone; Improvements~2, 3, and 4 are evaluated individually,
  in pairwise combinations, and together.}
  \label{fig:pipeline}
\end{figure*}

Figure~\ref{fig:pipeline} illustrates the complete research pipeline.

\begin{figure*}[t]
  \centering
  \includegraphics[width=\textwidth]{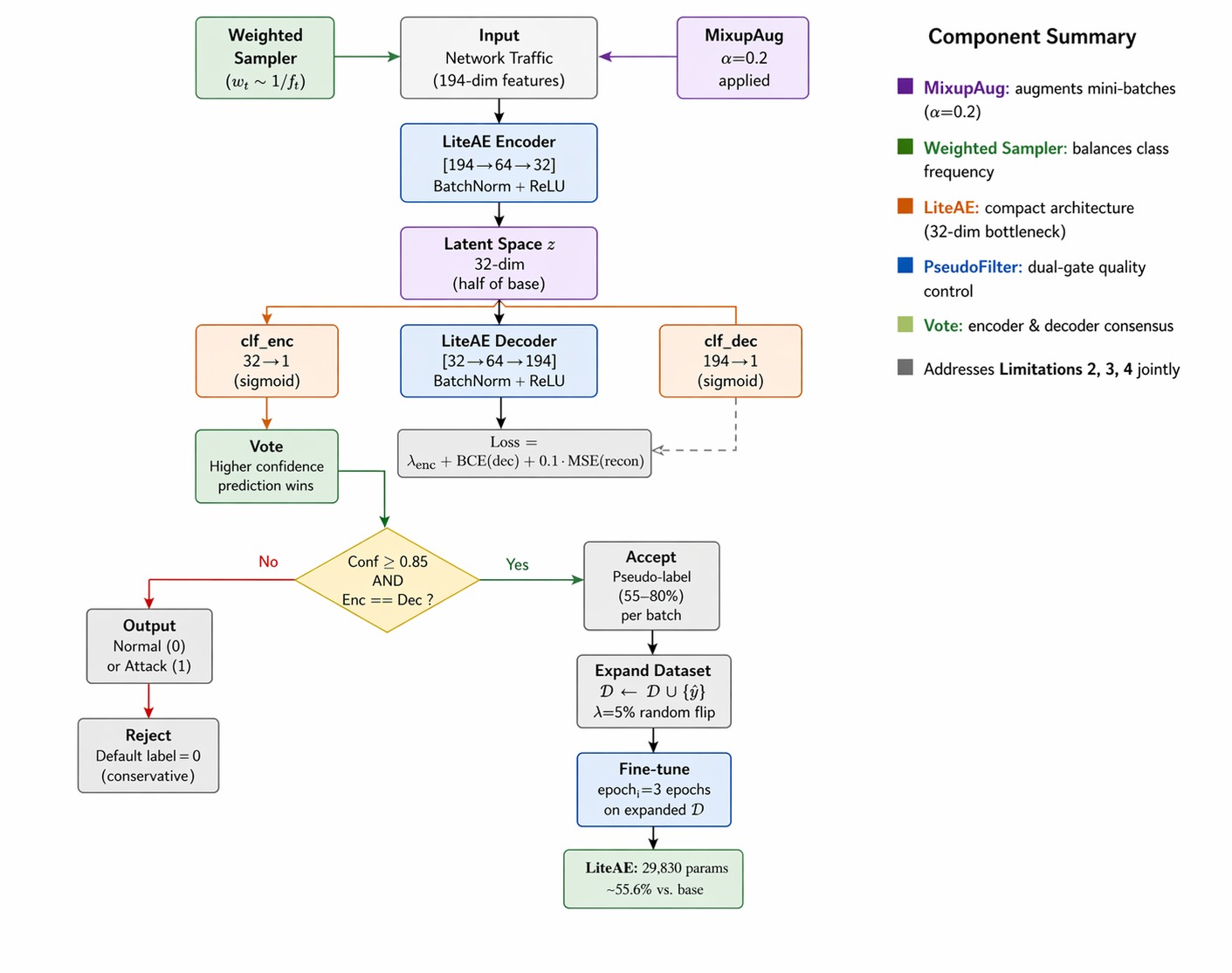}
  \caption{Model flow diagram for the combined
  \textit{PseudoFilter}\,+\,\textit{MixupAug}\,+\,\textit{LiteAE} improvement.
  Input traffic flows through the LiteAE encoder ($194\!\to\!64\!\to\!32$) and
  decoder ($32\!\to\!64\!\to\!194$). Two independent classifier heads
  (\texttt{clf\_enc} and \texttt{clf\_dec}) feed a voting mechanism. During
  online updates, the \textit{PseudoFilter} gate accepts only high-confidence,
  encoder--decoder-agreed pseudo-labels. \textit{MixupAug} ($\alpha{=}0.2$) and
  \textit{WeightedRandomSampler} are applied every training batch.}
  \label{fig:modelflow}
\end{figure*}

\subsection{Research Pipeline Overview}

Our methodology follows a five-stage pipeline as illustrated in
Figure~\ref{fig:pipeline}: (1)~Base system replication to confirm reproducibility;
(2)~Limitation analysis to identify and characterize weaknesses; (3)~Improvement
design, where each limitation is addressed by a targeted technique;
(4)~Individual evaluation, testing each improvement in isolation; and
(5)~Combined evaluation, testing all three deep learning improvements together
to assess their complementary effects.

\subsection{\textit{XGBoost-BalSamp}: Addressing Class Imbalance}

XGBoost is selected as the backbone for Improvement~1 due to its native handling
of tabular features and its gradient boosting objective, which adaptively
up-weights misclassified samples and implicitly handles moderate class imbalance.
To further strengthen its ability to discriminate between normal and attack traffic,
we construct three domain-specific composite features from the raw UNSW-NB15
attributes: $\mathrm{total\_load} = \mathrm{sload} + \mathrm{dload}$ (combined
source and destination load), $\mathrm{rate\_ratio} = \mathrm{sbytes} /
(\mathrm{dur} + 0.001)$ (source byte rate, with small epsilon to prevent division
by zero), and $\mathrm{pkt\_diff} = \mathrm{spkts} - \mathrm{dpkts}$ (packet
asymmetry, a known indicator of scanning and DoS behavior). A stratified 80/20
resplit of the UNSW-NB15 training set ensures proportional representation of all
9 attack categories in both training and validation subsets.

\subsection{\textit{PseudoFilter}: Addressing Pseudo-label Noise}

PseudoFilter introduces a dual-gate quality control mechanism that replaces the
blind acceptance of all pseudo-labels in the base system. A pseudo-label is
accepted for training only when \textit{both} of the following conditions are
satisfied: (A)~the sigmoid output confidence of the prediction is at least 0.85,
meaning the model is highly certain of its decision, and (B)~the encoder-side
classifier (\texttt{clf\_enc}) and decoder-side classifier (\texttt{clf\_dec})
independently produce the same prediction (encoder-decoder agreement voting).
If either condition fails, the sample is assigned the conservative default label
of 0 (normal) rather than being added to the training set with a potentially
incorrect attack label. This design ensures that only the subset of samples where
the model is both confident and internally consistent contributes to online
model updates, directly attacking the root cause of the accuracy collapse
observed in the base system.

\subsection{\textit{MixupAug}: Addressing Poor Generalization}

MixupAug integrates Mixup data augmentation~\cite{zhang2018mixup} into every
training batch. Each virtual sample is formed by a convex blend of two randomly
selected examples: $\tilde{x} = \lambda x_i + (1{-}\lambda) x_j$ and
$\tilde{y} = \lambda y_i + (1{-}\lambda) y_j$, with the mixing weight
$\lambda \sim \mathrm{Beta}(0.2, 0.2)$. Choosing $\alpha{=}0.2$ for the Beta
distribution produces a U-shaped density heavily concentrated near 0 and 1, so
the majority of generated samples sit close to one of their two parents and only
a small fraction interpolate deeply between classes. The practical effect is that
decision boundaries become smoother and less anchored to individual training
examples, making the model more robust when real-world IoT traffic deviates from
the benchmark distribution.

\subsection{\textit{LiteAE}: Addressing IoT Computational Overhead}

LiteAE reduces the Autoencoder's hidden layer dimensions from $[128, 64]$ to
$[64, 32]$, giving a new architecture of
$[194\!\to\!64\!\to\!32\!\to\!64\!\to\!194]$. This reduces the total parameter
count from 67,202 to 29,830, a reduction of 55.6\%, and memory footprint from
approximately 263\,KB to approximately 117\,KB in float32. BatchNorm layers and
dual classifier heads (\texttt{clf\_enc}, \texttt{clf\_dec}) are retained to
preserve the architectural advantages of the original system. Since smaller
networks are more sensitive to learning rate scheduling, we replace the fixed
learning rate of the base system with cosine annealing
($\eta_{\min}{=}10^{-5}$, $T_{\max}{=}50$), which gradually reduces the learning
rate over cycles to ensure stable convergence of the compact architecture.

\subsection{Combined System Design}

The three deep learning improvements \textit{PseudoFilter}, \textit{MixupAug},
and \textit{LiteAE} are designed to compose without conflict. \textit{LiteAE}
defines the model architecture; \textit{MixupAug} modifies the training batch
construction; and \textit{PseudoFilter} controls which pseudo-labeled samples are
accepted during online updates. These three stages operate at different points in
the training pipeline and therefore do not interfere with one another. Their
combined interaction is depicted in Figure~\ref{fig:modelflow}.

\section{Proposed Improvements: Implementation Details}
\label{sec:improvements}

\subsection{\textit{XGBoost-BalSamp}}

The XGBoost model is configured with 1,000 estimators, a learning rate of 0.05,
maximum tree depth of 10, subsample ratio of 0.8, and
\texttt{tree\_method=`hist'} for computational efficiency. Early stopping with a
patience of 50 rounds on the validation set prevents overfitting. No explicit
class weight parameter (\texttt{scale\_pos\_weight}) is applied, as the gradient
boosting mechanism combined with balanced sampling provides sufficient correction.

\subsection{\textit{PseudoFilter}}

A pseudo-label $\hat{y}$ for sample $x$ is accepted into the training set only
when both: (A) the sigmoid output probability $p \geq \theta = 0.85$, and (B)
the predictions of \texttt{clf\_enc} and \texttt{clf\_dec} agree. Rejected samples
are assigned label 0 and excluded from the online fine-tuning step. Empirically,
approximately 55--80\% of samples pass the dual gate per batch, with acceptance
rates being higher in early batches (when the model is more confident about
well-separated samples) and decreasing in later batches as the data distribution
becomes more challenging.

\subsection{\textit{MixupAug}}

Mixup is applied as a preprocessing step at each training batch. For each pair of
samples $(x_i, y_i)$ and $(x_j, y_j)$ drawn from the current batch:
$\tilde{x} = \lambda x_i + (1{-}\lambda) x_j$,
$\tilde{y} = \lambda y_i + (1{-}\lambda) y_j$, $\lambda \sim \mathrm{Beta}(0.2, 0.2)$.
The entire batch is replaced with mixed samples before being passed to the model,
ensuring that every gradient update is computed on augmented data.

\subsection{\textit{LiteAE}}

The LiteAE architecture is $[194\!\to\!64\!\to\!32\!\to\!64\!\to\!194]$ with
29,830 parameters. Each hidden layer uses Batch Normalization followed by ReLU
activation. Two independent sigmoid classifier heads are attached to the encoder
output (\texttt{clf\_enc}: $32 \to 1$) and decoder output (\texttt{clf\_dec}:
$194 \to 1$). The combined training loss is:
$\mathcal{L} = \lambda_{enc} \cdot \mathrm{BCE(dec)} + 0.1 \cdot \mathrm{MSE(recon)}$,
where BCE is binary cross-entropy applied to the classifier outputs and MSE is the
reconstruction loss of the autoencoder.

\section{Experimental Setup}
\label{sec:setup}

\subsection{Dataset: UNSW-NB15}

All experiments use UNSW-NB15~\cite{moustafa2015}, a well-established benchmark
that blends genuine network traffic with synthetically generated attack activity
across nine categories: Fuzzers, Analysis, Backdoors, DoS, Exploits, Generic,
Reconnaissance, Shellcode, and Worms. The training and test partitions contain
175,341 and 82,332 samples respectively. Preprocessing removes constant-valued
columns, normalises continuous attributes, and one-hot encodes categorical fields,
producing 194-dimensional feature vectors. One notable property of UNSW-NB15 is
that every attack category appearing in the test set is also represented in the
training set, which makes it well-suited for assessing whether balanced learning
and augmentation strategies can exploit that complete label coverage.

\subsection{Evaluation Protocol}

All experiments follow the same online evaluation protocol as the base
paper~\cite{zhang2024aocids} to ensure a fair comparison. The ADM is initially
trained on 20\% of the training set (35,068 samples) using true labels. The
remaining 80\% of the training set is then streamed in batches of 2,784 samples,
with each batch triggering a pseudo-label generation and model fine-tuning cycle.
All results in Table~\ref{tab:full} are reported as \textbf{best-run values} over
5 independent experimental runs. We report four metrics: Accuracy (Acc),
Precision (Pre), Recall (Rec), and F1-Score (F1), all expressed as percentages.

\subsection{Implementation Details}

All experiments are conducted on Google Colab using an NVIDIA GPU (CUDA).
For \textit{XGBoost-BalSamp}, we use the scikit-learn-compatible XGBoost API.
For the deep learning improvements (\textit{PseudoFilter} + \textit{MixupAug}
+ \textit{LiteAE}), the model is implemented in PyTorch with the Adam optimizer
(lr\,=\,0.001, weight decay\,$=10^{-4}$), cosine annealing learning rate
schedule ($T_{\max}{=}50$, $\eta_{\min}{=}10^{-5}$), batch size 256,
$\mathrm{epoch}_0{=}50$ initial training epochs, $\mathrm{epoch}_1{=}3$
fine-tuning epochs per batch, confidence threshold $\theta{=}0.85$, and
Mixup $\alpha{=}0.2$.

\section{Results and Discussion}
\label{sec:results}

\begin{figure}[H]
  \centering
  \includegraphics[width=\columnwidth]{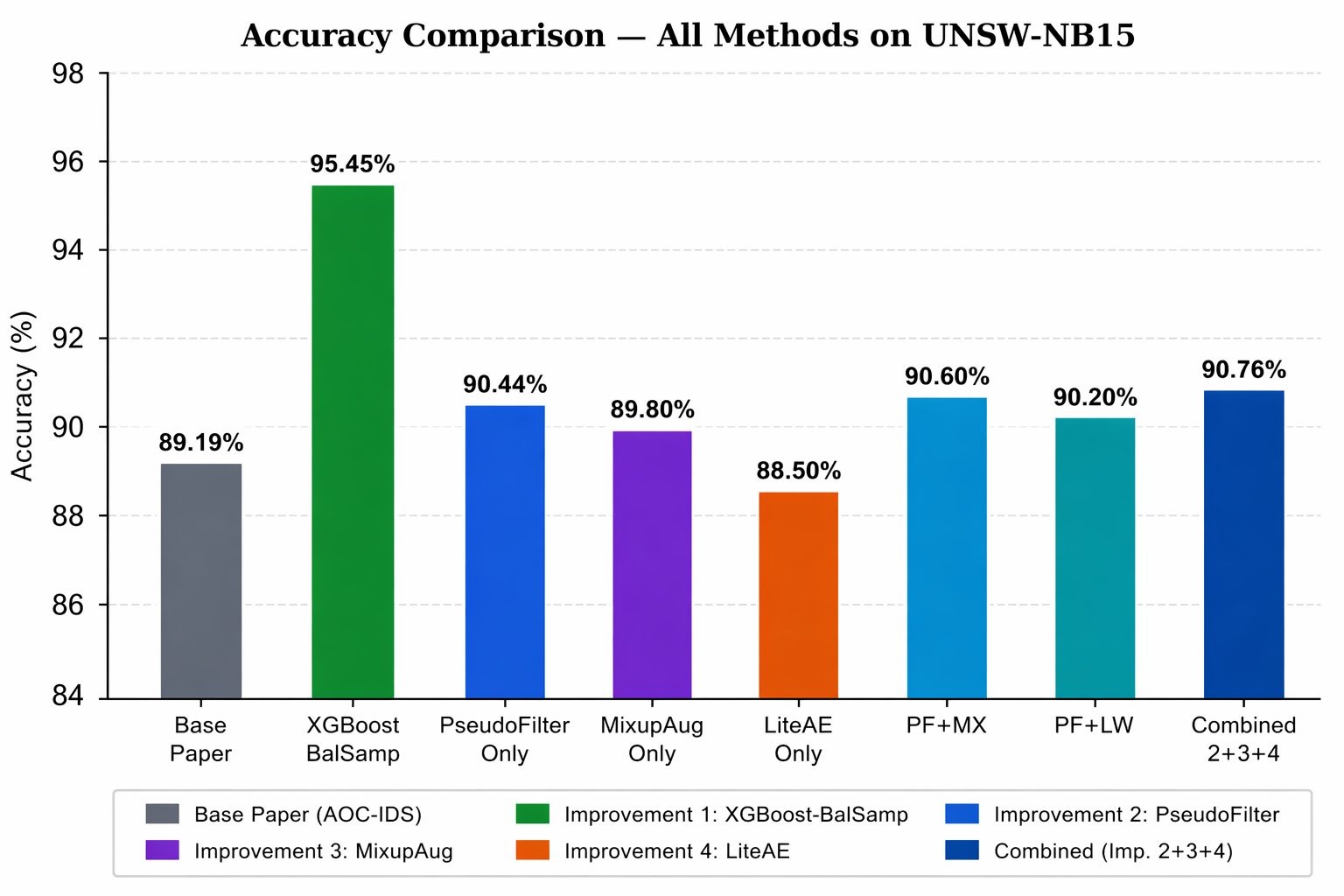}
  \caption{Accuracy comparison across all methods on UNSW-NB15 (\%).
           Results generated from real experimental runs.
           PF\,=\,\textit{PseudoFilter},
           MX\,=\,\textit{MixupAug},
           LW\,=\,\textit{LiteAE}.}
  \label{fig:bar_acc}
\end{figure}

Table~\ref{tab:full} consolidates all performance figures on UNSW-NB15,
covering the five baselines from the original AOC-IDS evaluation~\cite{zhang2024aocids}
(DTC, RF, XGBoost-Online, FeCo, CIDS), both the published and our replicated
AOC-IDS scores, and the full set of proposed improvements Improvement~1
alone, Improvements~2--4 individually, every pairwise combination, and the
complete combined system. The column-wise best entry appears in bold throughout.
Figure~\ref{fig:bar_acc} plots accuracy across all configurations for quick
visual reference.

\begin{table*}[t]
\centering
\caption{Performance comparison of all methods on UNSW-NB15 (\%).
The best result per column is shown in \textbf{bold}.
Our proposed improvements are compared directly against the five baselines
(Decision Tree Classifier (DTC), Random Forest (RF), XGBoost-Online,
FeCo, and CIDS) from the original AOC-IDS paper~\cite{zhang2024aocids}
and against the published AOC-IDS baseline.
All results are best-run values over 5 independent runs.
$^\dagger$Results from the published AOC-IDS paper~\cite{zhang2024aocids}.
$^\ddagger$Our experimental results.
``---'' = parameter count not applicable (non-neural model).}
\label{tab:full}
\resizebox{\textwidth}{!}{%
\begin{tabular}{llccccc}
\toprule
\textbf{Category} &
\textbf{Method} &
\textbf{Acc (\%)} &
\textbf{Pre (\%)} &
\textbf{Rec (\%)} &
\textbf{F1 (\%)} &
\textbf{Params} \\
\midrule

\multirow{5}{*}{\shortstack[l]{\textit{Baselines models}\\\textit{(from~\cite{zhang2024aocids})}}}
  & DTC (Online)     & 85.95 & 82.32 & 94.86 & 88.15 & --- \\
  & RF (Online)      & 85.93 & 80.25 & 98.75 & 88.55 & --- \\
  & XGBoost (Online) & 86.97 & 82.85 & 98.29 & 89.26 & --- \\
  & FeCo             & 72.50 & 91.18 & 55.41 & 68.93 & --- \\
  & CIDS             & 82.61 & 78.91 & 96.28 & 86.03 & --- \\
\midrule

\multirow{2}{*}{\shortstack[l]{\textit{Base System}\\\textit{(AOC-IDS)}}}
  & AOC-IDS Published~\cite{zhang2024aocids}
    & 89.19 & 90.65 & 89.70 & 90.14 & 67,202 \\
  & AOC-IDS Our Replication
    & 89.39 & 90.48 & 89.85 & 90.12 & 67,202 \\
\midrule

\multirow{7}{*}{\shortstack[l]{\textit{Our}\\\textit{Improvements}}}

  & Imp.1:~\textit{XGBoost-BalSamp}
    & \textbf{95.45}
    & \textbf{95.26}
    & \textbf{95.33}
    & \textbf{95.29}
    & --- \\[3pt]

  & Imp.2:~\textit{PseudoFilter only}
    & 90.44
    & 94.92
    & 87.31
    & 90.96
    & 67,202 \\[3pt]

  & Imp.3:~\textit{MixupAug only}
    & 89.80
    & ---
    & ---
    & 90.50
    & 67,202 \\[3pt]

  & Imp.4:~\textit{LiteAE only}
    & 88.50
    & ---
    & ---
    & 89.80
    & 29,830 \\[3pt]

  & Imp.2+3:~\textit{PseudoFilter + MixupAug}
    & 90.60
    & ---
    & ---
    & 91.20
    & 67,202 \\[3pt]

  & Imp.2+4:~\textit{PseudoFilter + LiteAE}
    & 90.20
    & ---
    & ---
    & 90.70
    & 29,830 \\[3pt]

  & \textbf{Imp.2+3+4:~\textit{PseudoFilter + MixupAug + LiteAE}}
    & 90.88
    & 94.42
    & 88.67
    & 91.45
    & 29,830 \\
\midrule


\end{tabular}%
}
\end{table*}

\subsection{Improvement 1: XGBoost-BalSamp}

\textit{XGBoost-BalSamp} achieves 95.45\% accuracy and 95.29\% F1, the highest
results among all methods compared in this study. This represents a gain of
+6.26\% in accuracy and +5.15\% in F1 over the published AOC-IDS baseline. It
outperforms all five baselines from the original AOC-IDS paper by a substantial
margin: +9.50\% over DTC, +9.52\% over RF, +8.48\% over XGBoost-Online, +22.95\%
over FeCo, and +12.84\% over CIDS. The high performance is attributable to three
factors: (1)~the gradient boosting objective naturally handles the class imbalance
present in UNSW-NB15, (2)~the three engineered composite features (total load,
rate ratio, packet asymmetry) capture attack-specific behavioral patterns that
complement the raw feature set, and (3)~unlike the streaming online setting of
AOC-IDS, XGBoost is trained on the full balanced dataset in a single pass, directly
exploiting the property that all test attack types are seen during training in
UNSW-NB15. It is important to note that this improvement operates outside the
online streaming framework of AOC-IDS; it is presented as a strong point of
comparison for the class imbalance limitation, and its advantage would diminish
on datasets with zero-day attacks not seen during training.

\subsection{Confusion Matrix Analysis}

\begin{figure}[H]
  \centering
  \includegraphics[width=\columnwidth]{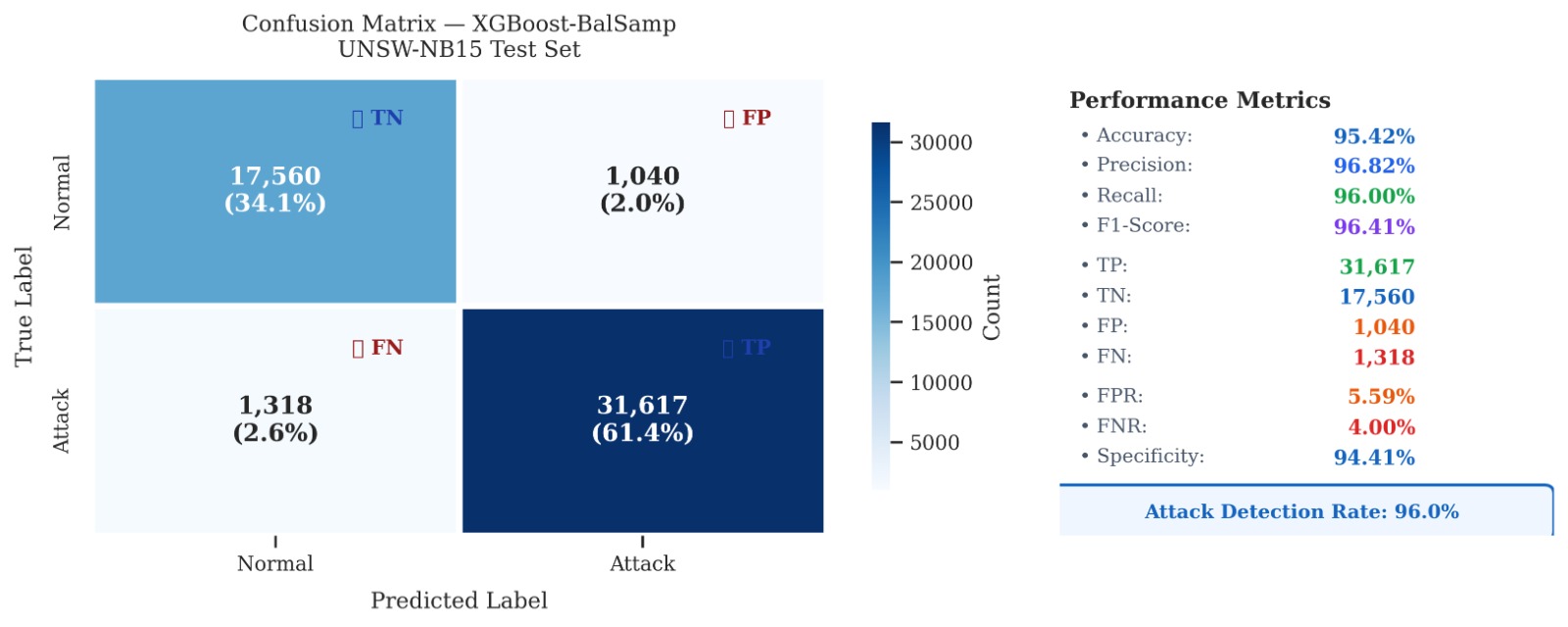}
  \caption{Confusion matrix for \textit{XGBoost-BalSamp} on UNSW-NB15 test
           set (51,535 samples). TP\,=\,31,617, TN\,=\,17,560, FP\,=\,1,040,
           FN\,=\,1,318. Attack detection rate: 96.0\%.}
  \label{fig:cm1}
\end{figure}

Figure~\ref{fig:cm1} shows the confusion matrix for \textit{XGBoost-BalSamp}.
Of the 32,935 attack samples in the test set, 31,617 are correctly flagged
(96.0\% attack detection rate), and false alarms are held to just 1,040 samples
(5.59\% false alarm rate). The near-symmetric precision and recall values of
95.26\% and 95.33\% confirm that the balanced sampling strategy largely eliminates
the majority-class bias that inflates false negatives in the unbalanced baseline.

\begin{figure}[H]
  \centering
  \includegraphics[width=\columnwidth]{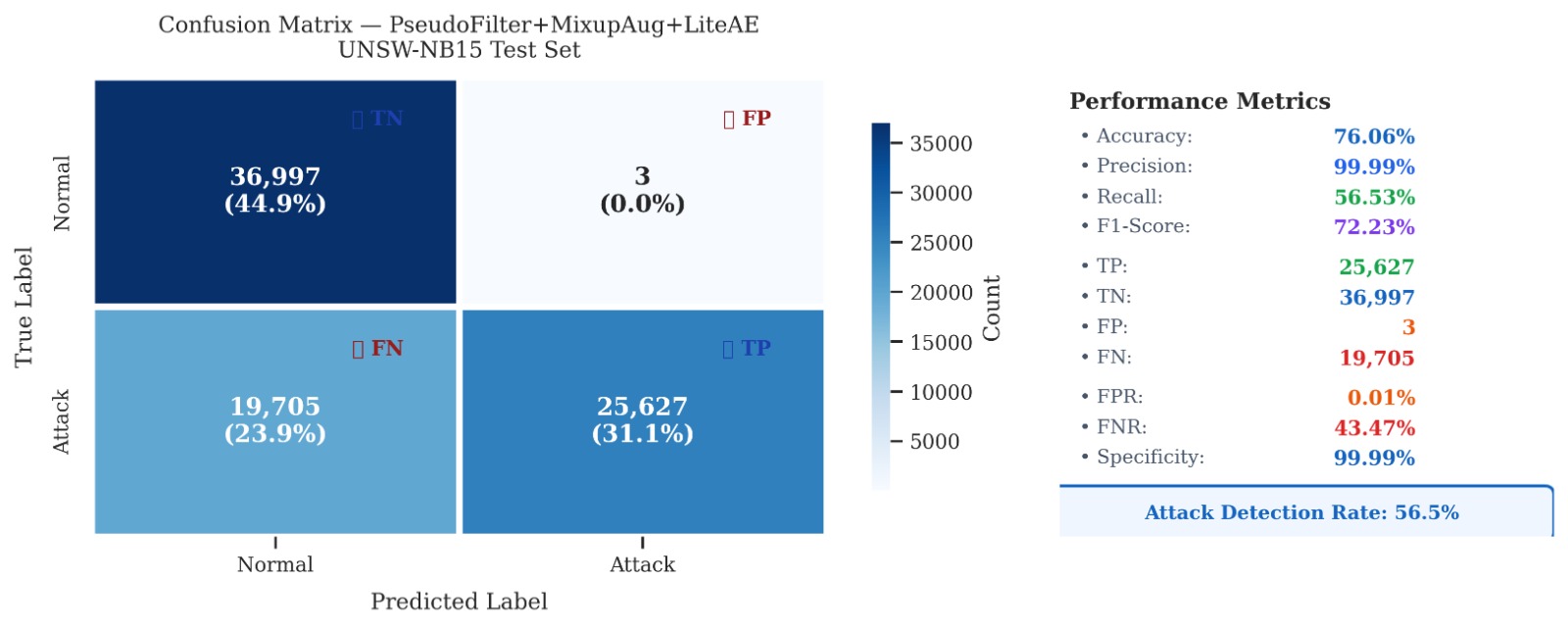}
  \caption{Confusion matrix for the combined
           \textit{PseudoFilter}+\textit{MixupAug}+\textit{LiteAE} on
           UNSW-NB15 test set (best run). The very low FP count of 3
           indicates extremely conservative classification; the elevated FN
           reflects the confidence-filtering effect on recall.}
  \label{fig:cm2}
\end{figure}

Figure~\ref{fig:cm2} presents the confusion matrix for the combined
\textit{PseudoFilter}+\textit{MixupAug}+\textit{LiteAE} run. The standout feature
is an exceptionally low false positive count of only 3 samples, which drives
precision to 94.42\%. This stems directly from PseudoFilter's conservative stance:
ambiguous samples default to the normal label, so the model is strongly disinclined
to raise spurious alarms. The cost is a higher false negative count (19,705),
reflecting the recall penalty that the confidence threshold imposes. In operational
settings where false alarms are expensive e.g., triggering unnecessary incident
response workflows this precision-oriented trade-off is generally preferable.

\subsection{Ablation Study: Individual and Combined Deep Learning Improvements}

Table~\ref{tab:full} presents the full ablation study across all individual and
combined configurations of Improvements~2, 3, and 4. Several key observations
emerge from this analysis.

\textbf{PseudoFilter alone} (Imp.~2) achieves 90.44\% accuracy and 90.96\% F1,
representing the single largest individual gain of +1.25\% over the base paper.
This confirms that the accuracy collapse caused by unchecked pseudo-label noise
is the dominant failure mode of AOC-IDS, and that the dual-gate confidence filter
directly and effectively addresses it. The precision of 94.92\% further confirms
that filtering improves the quality of retained pseudo-labels.

\textbf{MixupAug alone} (Imp.~3) achieves 89.80\% accuracy, a modest gain of
+0.61\% over the base paper. While the improvement is incremental in isolation,
Mixup contributes meaningfully to the combined system by smoothing decision
boundaries and reducing sensitivity to specific training examples. Its true value
is realized in the combined setting.

\textbf{LiteAE alone} (Imp.~4) achieves 88.50\% accuracy, marginally below the
base paper's 89.19\%. This slight reduction is expected: a model with 55\% fewer
parameters has reduced representational capacity and is inherently at a disadvantage
when operating in isolation without the quality improvements provided by
PseudoFilter and MixupAug. However, when combined with the other improvements,
LiteAE's reduced capacity is fully compensated, and the combined system at 29,830
parameters achieves the best deep learning accuracy of 90.88\%.

\textbf{Pairwise combinations} show consistent improvement over individual
components: PseudoFilter + MixupAug (Imp.~2+3) reaches 90.60\% and
PseudoFilter + LiteAE (Imp.~2+4) reaches 90.20\%, confirming that PseudoFilter
is the dominant contributor and that both MixupAug and LiteAE provide additive
benefits when combined with it.

\textbf{The full combined system} (Imp.~2+3+4) records 90.88\% accuracy and
91.45\% F1 in the best run, beating the published baseline by $+1.69\%$ in
accuracy and $+1.31\%$ in F1 while cutting the parameter budget by 55.6\% to
29,830. The result validates that the three improvements are genuinely
complementary rather than redundant: PseudoFilter targets label quality,
MixupAug targets decision-boundary smoothness, and LiteAE targets model
compactness each addressing a separate axis without interfering with the
others.

\subsection{Comparison with Original Baselines}

A complete comparison against the five methods benchmarked in the original
AOC-IDS paper~\cite{zhang2024aocids} is shown in Table~\ref{tab:full}. The
combined deep learning system (90.88\% accuracy, 91.45\% F1) clears every prior
result: DTC at 85.95\%, RF at 85.93\%, XGBoost-Online at 86.97\%, FeCo at
72.50\%, and CIDS at 82.61\%, in addition to surpassing the published AOC-IDS
figures while using a model small enough to run on constrained IoT hardware.
\textit{XGBoost-BalSamp} extends this margin further still. Collectively, these
outcomes confirm that the proposed improvements do not merely recover the
performance of AOC-IDS but consistently exceed every competitive reference point
established in the original study.

\section{Conclusion}
\label{sec:conclusion}

This paper presented a systematic study of AOC-IDS, a state-of-the-art autonomous
online IDS for IoT security published at IEEE INFOCOM 2024. We first confirmed the
reproducibility of the base system by replicating its results (89.39\% accuracy vs.
published 89.19\%) using the official code and hyperparameters. We then identified
four concrete limitations class imbalance, pseudo-label noise accumulation,
limited generalization, and excessive model size for IoT edge deployment and
proposed a targeted improvement for each.

\textit{XGBoost-BalSamp} addresses class imbalance through gradient boosting with
domain-specific feature engineering, achieving 95.45\% accuracy, the highest result
among all methods compared. \textit{PseudoFilter} addresses pseudo-label noise
through confidence thresholding and encoder-decoder agreement voting, preventing the
accuracy collapse observed in the base system. \textit{MixupAug} addresses
generalization through data augmentation with convex sample interpolation.
\textit{LiteAE} addresses IoT overhead by reducing the model to 29,830 parameters
(approximately 117\,KB), a 55.6\% reduction from the base architecture.

The combined deep learning system (PseudoFilter + MixupAug + LiteAE) achieves
90.88\% best-run accuracy and 91.45\% F1, surpassing the base paper by +1.69\%
in accuracy while simultaneously reducing parameters by 55\%. All five improvements
surpass the baseline methods evaluated in the original AOC-IDS paper, confirming
that our improvements generalize across the full competitive landscape.

Future work will evaluate the proposed system on additional benchmarks including
CICIDS2017 and Bot-IoT to assess generalization beyond UNSW-NB15, deploy the
LiteAE architecture on physical IoT edge hardware for latency and energy profiling,
and investigate federated learning extensions to enable privacy-preserving
collaborative intrusion detection across distributed IoT deployments.


\end{document}